\documentclass[a4paper,onecolumn]{agn6}
\usepackage{natbib}
\usepackage{graphicx}
\usepackage[a4paper]{hyperref}
\idline{}{1}
\begin{document}
   \title{Chandra discovery of activity in the quiescent nuclear black 
   hole of NGC821
}

   \author{A. Baldi \inst{1}, 
   G. Fabbiano \inst{1},
   S. Pellegrini \inst{2},
           A. Siemiginowska \inst{1},
	   M. Elvis \inst{1},
	   A. Zezas \inst{1},
          \and
          J.C. McDowell \inst{1}
}

   \institute{Harvard-Smithsonian Center for Astrophysics,
   60 Garden St, Cambridge, MA 02138, USA \email{abaldi@cfa.harvard.edu,
   gfabbiano@cfa.harvard.edu, aneta@cfa.harvard.edu, elvis@cfa.harvard.edu,
   azezas@cfa.harvard.edu, jcm@cfa.harvard.edu} 
         \and
             Dipartimento di Astronomia, Universit\'a di Bologna, via
Ranzani 1, 40127 Bologna, Italy \email{silvia.pellegrini@unibo.it}
             }

   \abstract{
We report the results of the {\it Chandra} ACIS-S observations of the
elliptical galaxy NGC~821, which harbors a supermassive nuclear black
hole (of $3.5 \times 10^7 M_{\odot}$), but does not show sign of AGN
activity.  A small, 8.5$^{\prime\prime}$ long ($\sim 1$~kpc at the
galaxy's distance of 23~Mpc), S-shaped, jet-like feature centered on
the nucleus is detected in the 38~ksec ACIS-S integrated exposure of
this region. The luminosity of this feature is $L_X \sim 2.6 \times
10^{39} \rm ergs~s^{-1}$ (0.3-10~keV), and its spectrum is hard
(described by a power-law of $\Gamma = 1.8^{+0.7}_{-0.6}$; or by
thermal emission with $kT >2$~keV).  We discuss two possibilities for
the origin of this feature: (1) a low-luminosity X-ray jet, or (2) a hot
shocked gas. In either case, it is a clear indication of nuclear
activity, detectable only in the X-ray band. Steady spherical accretion of
the mass losses from the central stellar cusp within the 
accretion radius, when coupled to a high radiative efficiency, already
provides a power source exceeding the observed radiative losses from
the nuclear region.

   }
   \authorrunning{A. Baldi et al.}
   \titlerunning{Chandra's discovery of nuclear activity in NGC821}
   \maketitle
%

\section{Introduction}


We currently know that virtually all galaxies host supermassive black holes
(SMBH) in their nuclei.  High resolution observations of the nuclei of
elliptical galaxies and bulges have established the presence of these
SMBHs (e.g. Magorrian et al 1998). 
However the lack of detectable nuclear emission in most galaxies 
where a SMBH is present makes the question more puzzling.  Although a
few faint AGNs have been detected in X-rays (e.g, in the $L_X/L_E \sim
10^{-(6-7)}$ range, where $L_E$ is the Eddington luminosity 
of the SMBH; Pellegrini et al. 2003), 
we still do not have a clear picture
of what impedes the formation of a luminous AGN.  

NGC~821 represents an example of such galaxies, hosting a SMBH and where
no nuclear emission has been detected so far.
It is an E6 galaxy located at a
distance of 23~Mpc, where the ACIS resolution
at the aim point corresponds to 55~pc. Prior to our {\it Chandra}
observation, NGC~821 had been observed, but not detected, in X-rays
with {\it ROSAT} ($ < 5 \times 10^{40} \rm erg~s^{-1}$; Beuing et
al. 1999). This limit is $\sim 10^5$ times below the Eddington luminosity
of the nucleus, based on the SMBH mass of $2.8-5.8 \times 10^7
M_{\odot}$ (Gebhardt et al. 2003). As
reported in Ho et al. (2003), NGC~821 has not been
detected in radio continuum nor in optical emission lines (H$\alpha$
and H$\beta$), and is a good example of quiescent SMBH. 
In this talk we report the results of the {\it Chandra} 
ACIS observation of NGC~821, that has led
to the discovery of an S-shaped feature, suggestive of either a weak
two-sided X-ray nuclear jet, or of hot shocked gas.

\section{Observations and Data Analysis}

NGC~821 was observed with {\em Chandra} ACIS-S
(PI: Fabbiano) on November 26, 2002 (ObsID: 4408) and on December 1,
2002 (ObsID: 4006) for a total exposure time of 38~ks.
Table~\ref{obslog} is a summary of the relevant properties of
NGC~821 and of the observing log.

\begin{table}
\tiny 
\begin{center}
\begin{tabular}{lcccccccc}
\hline
M$^0_{B_T}$ & D & Diam.(') & N$_H$ & L$_X$(erg~s$^{-1}$) & 
$M_{\bullet}$ & ObsID & Date & T$_{exp.}$\\
(mag) & (Mpc) &  $\sigma_o$(km~s$^{-1}$) & (cm$^{-2}$) & 
$L_{H\alpha}$(erg~s$^{-1}$) & ($10^7 M_{\odot}$) & & & (ks)\\
\hline
-20.71&23&2.6&6.2$\times 10^{20}$&$<$5$\times 10^{40,}$\footnotemark
&2.8-5.8\footnotemark&4408&Nov. 26, 2002& 24.6\\
$...$ &$...$ &209\footnotemark &$...$ &
$< 1.3 \times 10^{38,}$\footnotemark &$...$ &4006& Dec. 1, 2002&13.3\\
\hline

\end{tabular}
\end{center}
\footnotetext{}{$^1$Beuing et al. 1999;}
\footnotetext{}{$^2$Gebhardt et al. 2003, rescaled for the distance adopted here;}
\footnotetext{}{$^3$Prugniel \& Simien 1996;}
\footnotetext{}{$^4$Ho et al. 2003.}

\caption{NGC~821: Properties and {\it Chandra} ACIS-S Observation Log.
Unless otherwise noted, the galaxy properties are as listed in NED 
(NASA / IPAC Extragalactic Database).\label{obslog}
}
\end{table}

\subsection{X-ray image}

From the dataset obtained merging the two observations, images were extracted in
three spectral bands (Red = 0.5 - 1~keV; Green = 1 - 2~keV; Blue = 2 -
4~keV).

A high resolution `true-color' image of the central region of
NGC~821 is shown in fig.~\ref{fig1}a, where the data are displayed at the
original observed resolution, without smoothing. This figure shows
clearly a north-south elongated, hard central feature, centered
on the nucleus of NGC~821, at RA=$02^h 08^m 21.14^s$,
Dec=$+10^o 59^\prime 41.7^{\prime\prime}$ (J2000, with uncertainty
of $1.25^{\prime\prime}$; from the 2MASS survey, as reported in NED). 
The general form is suggestive of a two-sided bent
jet, or S-shaped filament centered on the nucleus. 
This feature is approximately 8.5'' long, corresponding to $\sim
1$~kpc at the distance of NGC~821.

\subsection{Spatial analysis of the nuclear
feature}

   \begin{figure}
   \centering
  \caption{Left: True color image of the central region of NGC~821, unsmoothed. 
The cross and surrounding circle represent the 2MASS nuclear position and 
uncertainty, from
NED. Right: A larger 
field image  showing both the $wavdetect$ source regions (yellow) and the spectral 
extraction regions for the `jet' and the diffuse emission (light 
blue).} 
     \label{fig1}
\end{figure}
%

In the central region shown in fig.~\ref{fig1}a, there are four sources: the
isolated point-like source at the north-east of the central complex
(source NE),
and three sources in the central elongated emission region,
identified by ellipses in fig.~\ref{fig1}b, and named S1, S2 and S3, from
north to south. To establish their spatial properties, we have
compared their spatial distribution of counts with
that of the on-axis image of the quasar GB~1508+5714 (Siemiginowska et
al 2003a), which can be used as a good representation of the
{\it Chandra} ACIS-S PSF. With a
count rate of $\sim 0.05 \rm~count~s^{-1}$, the image of GB~1508+5714 (ObsID 2241)
is not affected by pile-up, and contains 5,300 counts within 2'' of the
centroid of the count distribution. From this image, we determine the
ratio of counts within the 1"-2" annulus to those in the central 1"
radius circle to be {\it Ratio}(PSF)$ = 0.043 \pm0.001~(1\sigma)$. The isolated 
NE source in
this central field yields {\it Ratio}(NE)$ = 0.057 \pm 0.054$ (from a total
of 36 source counts), entirely consistent with that of our reference
quasar, confirming that GB~1508+5714 gives a good representation of the PSF. 
Instead, the
analogous ratios for the background-subtracted counts from S1, S2 and
S3 demonstrate that the emission is extended in all cases.
Using the {\it wavdetect} centroids, we obtain {\it Ratio}(S1)$ =  0.81 \pm 0.26$,
{\it Ratio}(S2)$ =  0.94 \pm 0.27$, and {\it Ratio}(S3)$ =  0.29 \pm 0.10$. The
total number of source counts in the three cases are 56, 67, and 71,
respectively, larger than for source NE.  This comparison demonstrates that the 
spatial
distributions of the source counts from S1 and S2 are definitely not
consistent with the PSF. 
Therefore, these three central emission regions are clearly not point-like and 
cannot be explained with the serendipitous positioning of three luminous
galaxian LMXBs in NGC~821.

Given that the central emission (S2) is not consistent with a
point-source, we can only estimate a 3~$\sigma$ upper limit on the
luminosity of a nuclear point-like AGN. A $1" \times 1"$ ($2
\times 2$ pixels) sliding cell over the entire area covered by the
S-shaped feature (assuming as the background level the maximum value detected
in the sliding cell) yields a  0.3-10~keV 
$L_X < 4.2 \times 10^{38} \rm erg~s^{-1}$ for a $\Gamma =1.8$ power-law spectrum
and Galactic $N_H$. This limit indicates 
that a central point-like AGN would have a luminosity not exceeding 
that of normal LMXBs. 

\subsection{Spectral analysis}

\begin{figure}
   \centering
   \resizebox{0.8\hsize}{!}{\includegraphics{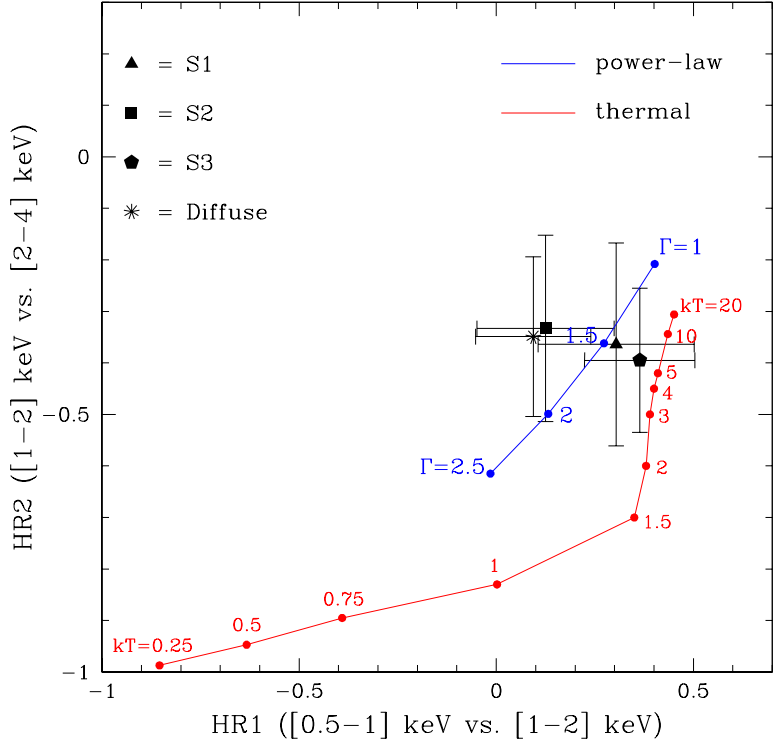}
   }
   \caption{X-ray colour-colour diagram of `jet' and diffuse emission regions. Typical
hardness ratios for a power-law model ($\Gamma=1-2.5$) are plotted in blue. Typical
hardness ratios for a thermal model ($kT=0.25-20$ keV) are plotted in red. Galactic
line-of-sight absorption ($N_H = 6.4 \times 10^{20} \rm cm^{-2}$)
is assumed in both models.} 
     \label{fig2}
\end{figure}

Hardness ratios (HR1=M-S/M+S; HR2=H-M/H+M; where S=0.5-1 keV, M=1-2
keV, and H=2-4 keV) are plotted (with 1~$\sigma$ statistical errors)
in fig.~\ref{fig2}, and compared with power-law and Raymond-Smith emission
models. The hardness ratios of the S1, S2, S3 regions are all
consistent with hard emission, either a power-law spectrum with $\Gamma$ between 
$\sim1$ and
$\sim2.2$, or thermal emission with $kT \sim 2-20$~keV. 
The diffuse X-ray emission of an elliptical galaxy is the
combination of the soft emission of the hot ISM and of the hard emission of
the population
of LMXBs below our detection threshold (see e.g, Kim \& Fabbiano
2003). In NGC~821 the hardness ratios of the diffuse emission 
are hard, suggesting a dominant LMXB component and little hot ISM.

A proper spectral analysis was possible only on the whole S-shaped feature and
not on the individual S1, S2 and S3 sources.
Therefore spectra were extracted both from the S-shaped feature
(regions in fig.~\ref{fig1}b) and the surrounding diffuse emission (within a 
20$^{\prime\prime}$ radius)

With XSPEC, we fitted the data in the 0.3-10~keV energy range,
rebinned at $N>15$ counts per energy bin. For the S-shaped 
feature we adopted an absorbed power-law model (XSPEC model: {\em
wabs(wabs(pow))}), with $N_H$ consisting of both a Galactic and an
intrinsic component.
The results of the fits are listed in Table~\ref{fitres}, with 90\%
errors on one significant parameter. The  S-shaped  emission is well fitted
with a power-law spectrum with $\Gamma \sim 1.8$, typical of AGN
spectra, although the uncertainties are large.

\begin{table}
\begin{center}
\begin{tabular}{ccccccc}
\hline
 & Net Counts & $\chi^2/dof$ & $N_H$ &
kT & $\Gamma$ & $L_X (0.3-10~keV)$\\
 & (0.3-10 keV) & & ($\times10^{21}$ cm$^{-2}$) & (keV) & & (erg~s$^{-1}$)\\
\hline
S-shaped & 141 & 6.7/5 & $1.41_{-1.41}^{+1.84}$ & $...$ & $1.82_{-0.56}^{+0.71}$ & 
$2.6\pm0.5\times10^{39}$\\
$...$ & $...$ & 6.5/5 & $ 0.75_{-0.75}^{+2.34}$ & $5.14_{-2.99}^{+50.00}$& ...& 
$1.9_{-0.4}^{+1.4} \times 10^{39}$\\
Diffuse & 174 & 6.5/10 & $<9.00$ & $0.46_{-0.25}^{+0.33}$ & $1.27_{-0.68}^{+1.13}$ & 
$3.5\pm0.6\times10^{39}$\\
Thermal &$...$ &$...$&$...$&$...$&$...$& $3.4\pm0.1\times10^{38}$\\
\hline
\end{tabular}
\end{center}
\caption{Results of Spectral Fits \label{fitres}}
\end{table}

For the diffuse emission, following e.g. Kim \&
Fabbiano (2003), an optically thin thermal component ($Z=0.3$) was 
added to the power-law model (XSPEC model: {\em wabs(wabs(apec+pow))}).
The diffuse emission spectrum (see Table~\ref{fitres})
is consistent with a hot gas with
a temperature $kT\sim 0.5$~keV, or cooler, typical of X-ray-faint
early-type galaxy halos (e.g. Pellegrini \& Fabbiano 1994). 
The power-law component is
needed to obtain an acceptable fit for the diffuse emission, as clearly
suggested by fig.~\ref{fig2}, 
in agreement with the presence of an unresolved LMXB population.

Table~\ref{fitres} also lists the best-fit unabsorbed
luminosities for the S-shaped feature, the total diffuse emission and the thermal 
component of the diffuse emission. The latter is only 10\% of the total diffuse 
emission, indicating that NGC~821 is singularly devoided of 
hot ISM.

More details on data reduction and analysis can be found in Fabbiano et al. (2004).

\section{Discussion}

Our observations fail to detect a
point-like source at the nucleus, down to a $3 \sigma$ limit of
$L_X(0.3-10~keV) < 4.2 \times 10^{38} \rm erg~s^{-1}$, $\sim 100$
times fainter than the ROSAT limit (see Table~1). The nucleus is not
detected in the FIR or in H$_2$ (Georgakakis et al. 2001) arguing
against a strongly obscured AGN (and against a nuclear
starburst). There is also no sign of strong intrinsic absorbing column
in the X-rays (see Table~2). The general AGN `quiescent state' is also
supported by the lack of radio continuum and optical line emission (Ho et al. 2003).  
We detect instead an elongated ($\sim1$kpc), 
possibly bent, emission feature, strongly suggestive of 
a two-sided X-ray jet or S-shaped filament, with a
hard spectrum consistent with a $\Gamma \sim 1.8$ power-law 
or, if thermal, $kT>2$~keV. The
X-ray luminosity of this feature is $\sim 1.9-2.6 \times 10^{39} \rm
ergs~s^{-1}$, corresponding to $\sim 5 \times 10^{-7}$ of the
Eddington luminosity of the SMBH.

\subsection{Is a jet directly emitting the hard X-rays?}

The S-shaped, hard emission centered on the nucleus of NGC~821 
could be a two-sided nuclear jet, as in
radio galaxies. The spectral power-law slope of
this emission has large uncertainties (Table~2), but is consistent
with the X-ray spectra of other jets (Siemiginowska et al. 2003b; Sambruna et al 2004). 
However, unlike other extragalactic X-ray jets seen in luminous
AGNs, where $L_X(jet)/L_X(AGN)\sim $ 1--15\%, this `jet' 
has no associated core X-ray source, implying
$L_X (jet) / L_X (AGN) > 6$. 
However, 
the NGC~821 `jet' could be similar to the somewhat steeper spectrum
($\Gamma\sim$2.3) M87 jet (Wilson \& Yang 2002), where the nuclear
point-like AGN is fainter compared with the jet ($L_X (jet) /
L_X (AGN)\sim 2$).  The 0.5~kpc
(one-sided) scale of the NGC~821 `jet' is also similar to the
$\sim$1.5~kpc of the M~87 jet, while most radio/X-ray jets extend over
much larger distances, up to 300~kpc (Siemiginowska et al. 2003b). 

Synchrotron emission is plausible to be the emission mechanism 
responsible for the S-shaped emission of NGC~821, in the jet hypothesis: 
$\alpha_{radio-X} \leq 0.7$ (estimated from the radio
flux limit of 0.5~mJy at 5~GHz and the X-ray flux at 1~keV);  this
$\alpha_{radio-X}$ is consistent with the X-ray slope, and is typical
of the synchrotron slope observed in radio lobes (Peterson 1997).
A radio detection not far
below the current limit is expected if synchrotron radiation produces
the S-shaped emission at the center of NGC~821. 

However $\alpha_{radio-X} \leq 0.7$ is also consistent with the values
of $\alpha_{radio-X}$=0.7-0.8 reported for knots in powerful jets by
Sambruna et al. (2004) who favor an Inverse Compton origin for the
X-ray flux for most of the jet knots, based on the X-ray fluxes lying
above an extrapolation of the radio-optical slope ($\alpha_{RO} >
\alpha_{OX}$). The seed photons for
Comptonization could be either from the synchrotron photons within the
jet (the `self-Compton' case) or could be the external to the jet photon field.
In the self-Compton process the ratio of the synchrotron to Inverse
Compton luminosities is given by the ratio of energy densities
of the magnetic to the synchrotron radiation field. 
In equipartition both luminosities are of the same order, and since
the observed X-ray luminosity is a factor of at least 10$^3$ higher than
the radio upper limit, the self-Compton case is excluded.

External Comptonization is also ruled out: Felten \& Morrison (1966,
eq.47) showed that $I_S/I_C = U_B/U_{ph}(\nu_C/\nu_S)^{(3-m)/2}$. Here
$I_S$ and $I_C$ are intensities of synchrotron and inverse Compton
emission, $U_B$ is the energy density of magnetic field, $U_{ph}$ the
energy density of the external photon field, $\nu_C/\nu_S$ is the
ratio between the frequency of the Compton scattered photon and the
frequency of the synchrotron photon, $m$ is the power law index of the
electron distribution, which is linked to the spectral index,
$\alpha_s$ of the synchrotron emission so $m=1-2\alpha_s$. The quantities
above can be easily measured from the X-ray luminosity, the starlight from
the central cusp of NGC~821 (Gebhardt et al. 2003), and the
0.5~mJy flux limit at 5~GHz.
Considering the radial dependence of the optical photon field,
we obtain a maximum
predicted Inverse Compton emission due to scattering of the starlight
ranging from $3.7 \times 10^{36} \rm erg~s^{-1}$
at 1~pc galactocentric radius, to 
$7.9 \times 10^{35}  \rm erg~s^{-1}$ at 500~pc (the maximum jet extension). 
We conclude that Inverse Compton radiation 
would be a small contribution to the X-ray emission, suggesting 
that synchrotron may be the dominant emission mechanism if the S-shape feature is
indeed a jet.

\subsection{Is hot gas responsible for the hard emission?}

If the origin of the hard S-shaped emission is thermal, the most likely scenario, 
suggested by analogies with another galaxy, leads to 
the presence of shocks in the ISM, resulting from an
outburst of nuclear activity, as in the case of NGC~4636 (Jones
et al 2002).
The hot `arms' of this galaxy, 
a giant elliptical in Virgo with no reported nuclear activity or jets, suggest
an interesting analogy with the S-shaped feature of NGC~821. These arms, 
having a larger spatial scale (8~kpc) and a lower
temperature ($\sim 0.5 - 0.7$~keV) than the S-shaped feature of
NGC~821, are two symmetric features crossing the
galaxy center, discovered in the {\em Chandra} ACIS data of NGC~4636. The 
NGC~4636 arms are accompanied by a temperature increase with respect to the 
surrounding hot ISM,  which led to the suggestion by Jones et al. of shock 
heating of the ISM caused by a nuclear outburst. 
As discussed in \S~2.3, the S-shaped feature of NGC~821 could similarly be hotter than 
its surroundings. Assuming a temperature of kT=3~keV for this feature, close to the
lower limit suggested by our spectral fit (Table~2), we 
obtain a density $ n = 8.59_{-1.07}^{+1.06} ~\rm cm^{-3}$ (errors at 90\%).
Using the best-fit value (and a temperature of 3~keV), we obtain 
a thermal pressure exceeding by a factor of $\sim 14$ that quoted in \S 3.1 for the
surrounding hot ISM, suggesting a non-equilibrium situation, if only thermal
pressures are involved. However, given the uncertainties in both $T$ and $n$, the 
thermal pressures of both S-shaped feature and surrounding ISM could be similar.
It is clear that deeper {\it Chandra} observations are needed to better constrain the
energetics of this feature.


\subsection{Is accretion present?}

Whether it is a jet or hot shocked gas, the S-shaped feature of
NGC~821 requires a considerable energy input, an obvious source of
which is accretion onto the nuclear SMBH.  Taking at face value the
indication of the two-component fit of the circum-nuclear diffuse
emission, which is consistent with the presence of a $kT\sim 0.5$ keV
thermal component, we can estimate the nuclear accretion rate, in the
steady spherical accretion scenario of Bondi (1952). This estimate is
rough, because the gravitational capture radius (which depends on the
gas temperature and on the mass of the SMBH; see the textbook by Frank,
King \& Raine 2002) is $r_{acc}\sim 3-23$ pc in our case, smaller than
the physical resolution of the image (55~pc, Table~1). 
Based on the circum-nuclear gas
temperature and density, that we estimate from the emission
measure of the diffuse thermal component to be $n = 4.1_{-1.5}^{+10.9} \times
10^{-3} \rm cm^{-3}$, the Bondi mass accretion
rate (also following Frank, King \& Raine 2002) is $\dot M
_{acc}~=1.1~\times~10^{-7} - 2.0~\times~10^{-5}~\rm M_{\odot}~\rm
yr^{-1}$, including again all the uncertainties in the SMBH mass, $kT$ and
$n$. The corresponding luminosity is $L_{acc}~=~6.2 \times 10^{38} -
1.1 \times 10^{41}\, \rm ergs~s^{-1}$, with the customary assumption of
10\% accretion efficiency.  
The luminosity of the S-shaped feature is within this
range, therefore in principle it could be explained by Bondi accretion
of the hot ISM. 

However, since the hot ISM is the thermalized integrated result of the stellar
mass losses, as a minimum one expects the total stellar mass loss rate
$\dot M_{\star}$ within $r_{acc}$ to be accreted (there may also be gas
inflowing from outside $r_{acc}$). $\dot M_{\star}$ can easily be
obtained from the luminosity density profile recovered from HST data
for the central galaxy region (Gebhardt et al. 2003). Using
a conversion factor from luminosity to mass loss rate for an old
stellar population at the present epoch (e.g., Ciotti et al. 1991),
this leads to $\dot M_{\star} = 9.8\times 10
^{-6}$ and $2.6\times 10^{-4}$~M$_{\odot}$yr$^{-1}$,  for the two
extreme values of $r_{acc}$. These $\dot M_{\star}$ values are
respectively a factor of $\sim$90 and $\sim$13 larger than the $\dot
M_{acc}$ values derived above for the same $r_{acc}$. Given that this
estimate of $\dot M_{\star}$ is quite robust, we must
conclude that either the derivation of $\dot M_{acc}$ above is
inaccurate, or the gas is not steadily inflowing within $r_{acc}$.
The former possibility cannot be excluded with the present data, since in our 
calculation of  $\dot M_{acc}$ we used a density $n$ value that is an
average measured over a region extending much farther out than $r_{acc}$;
$n(r_{acc})$ is likely to be significantly higher than this average 
(e.g., a factor of $\sim 30 $ times higher, for a
$n\propto r^{-0.9}$ profile). 

Assuming that it is just the stellar mass loss rate
within $r_{acc}$ that is steadily accreted, accretion luminosities
$>20$ times larger than the observed $L_X$ of the hard emission are
recovered (from $L_{acc}=0.1\dot M_{\star}c^2$). Therefore 
the possible scenarios for NGC~821 are: (a) accretion occurs but with low
radiative efficiency; (b) accretion sustains a
jet (this can be coupled again to a low radiative efficiency), whose
total power is of the order of $L_{acc}$, as in the modeling for
IC~1459 (Fabbiano et al. 2003) and M87 (Di Matteo et al. 2003); (c)
accretion is unsteady and therefore the hot ISM in the nuclear region
needs not be inflowing (Siemiginowska et al 1996; Janiuk et al 2004). 
In this case the feedback from the central
SMBH can be either radiative (Ciotti \& Ostriker 2001) or mechanical
(Omma et al. 2004) and make accretion undergo
activity cycles: while active, the central engine heats the
surrounding ISM, so that radiative cooling -- and accretion -- are
offset; then the central engine turns off, until the ISM starts
cooling again and accretion resumes.  NGC~821 may be in a stage of
such a cycle when a nuclear outburst has recently occurred. Note
that the accretion luminosity that is radiatively absorbed by the
ISM during an outburst (Ciotti \& Ostriker 2001) largely exceeds 
the hard thermal emission observed at the center of NGC~821.

The presence of the central S-shaped feature that is
{\em hotter} than the surrounding gas is uncontroversial evidence that
central heating is at work, and therefore some type of feedback from
the SMBH is occurring. The unsteady scenario seems promising to fit
adequately the case of NGC821, also because this galaxy is clearly hot gas poor,
as if recently swept by an outburst-driven wind.  
From our ACIS-S data we estimate an upper limit of $\sim 4
\times 10^6 M_{\odot}$ on the amount of hot ISM, many orders of
magnitude smaller than for hot gas rich ellipticals (see Fabbiano 1989).

\begin{acknowledgements}
We thank the {\it Chandra} X-ray Center DS and SDS teams for their efforts in reducing 
the data and developing the software used for the data reduction (SDP) and
analysis (CIAO). We have used the NASA funded services NED and ADS, and browsed 
the Hubble archive.
This work was supported by NASA contract NAS~8--39073 (CXC) and NASA grant GO3-4133X\end{acknowledgements}

\bibliographystyle{aa}

\end{document}